\documentclass[letterpaper,twocolumn,10pt]{article}
\usepackage{usenix}

\usepackage{tikz}
\usepackage{amsmath,amssymb}
\usepackage{algorithm}
\usepackage[noend]{algpseudocode}
\usepackage{xspace}
\usepackage{booktabs}
\usepackage{subfig}
\usepackage{algorithmicx}

\newcommand{\sys}{\textsc{Nebula}\xspace}
\newcommand{\phm}[1]{\vspace{.4em}\noindent\textbf{#1}\hspace{.5em}}

\newenvironment{componentlist}{%
    \begin{list}{\textbullet}{%
        \setlength{\leftmargin}{1em}%
        \setlength{\labelwidth}{0.5em}%
        \setlength{\labelsep}{0.5em}%
        \setlength{\topsep}{2pt}%
        \setlength{\partopsep}{0pt}%
        \setlength{\itemsep}{1pt}%
        \setlength{\parsep}{0pt}%
    }%
}{%
    \end{list}%
}

\begin{document}

    \date{}

    \title{
        \sys: Distributed Position-Independent Caching for LLM Serving
    }

    \author{
        {\rm Yifei Liu\textsuperscript{1}\thanks{%
        \protect\hypertarget{Hfootnote.\theHfootnote}{}%
        \protect\hypertarget{fn:equal}{}Both authors contributed equally to this work.}\hspace{.4em}%
        \thanks{\protect\hypertarget{Hfootnote.\theHfootnote}{}%
        Yifei Liu is the corresponding author (liuyifei5@xiaohongshu.com).}
        \quad Junbin Pan\textsuperscript{1}\hyperlink{fn:equal}{\footnotemark[1]}
        \quad Junhao Hu\textsuperscript{3}
        \quad Yang Liu\textsuperscript{2}
        \quad Junyi Chen\textsuperscript{2}}\\
        {\rm Juntong Wu\textsuperscript{3}
        \quad Chen Chen\textsuperscript{2}
        \quad Haiyuan Zhao\textsuperscript{1}
        \quad Xiaoxu Chen\textsuperscript{1}
        \quad Weihang Chen\textsuperscript{1}}\\
        {\textsuperscript{1}Xiaohongshu
        \quad \textsuperscript{2}Shanghai Jiao Tong University
        \quad \textsuperscript{3}Peking University}
    }

    \maketitle

    \begin{abstract}
Modern LLM workflows, such as retrieval-augmented generation (RAG) and LLM agents, build long prompts from multiple independent text segments, making prefill a major source of computation.
Reusing KV caches is therefore important for reducing serving costs and latency: distributed caching systems spread KV cache across a cluster to increase aggregate capacity, while position-independent caching (PIC) enables reuse of recurring segments under different prefixes.
However, existing systems distribute KV caches but still rely on prefix caching (PC), while existing PIC methods enable position-independent reuse without cluster-wide resource management.

We present \sys, the first distributed LLM serving framework to treat PIC segments as first-class citizens, enabling remote PIC reuse and parallel prefill.
To track PIC's fine-grained, asynchronous state changes, \sys combines optimistic accounting with event-driven updates to provide a timely and truthful cluster view for cache management and prefill scheduling.
To accommodate diverse reuse patterns across workloads, \sys leverages cluster-wide ghost-hit feedback to automatically tune the allocation of shared cache capacity between PC and PIC.
Finally, \sys incorporates queueing, computation, and KV transfer costs into a unified prefill completion time model to guide elastic parallelism adjustment and locality-aware compute placement.
Across four models and four real-world RAG and agentic workloads, \sys reduces median TTFT by an average of 61.4\% compared with the strongest existing baseline and recovers an average of 76.1\% of the tokens missed by prefix caching.
\end{abstract}

    \section{Introduction}
\label{sec:intro}

\begin{figure}[t]
    \centering
    \includegraphics[width=\linewidth]{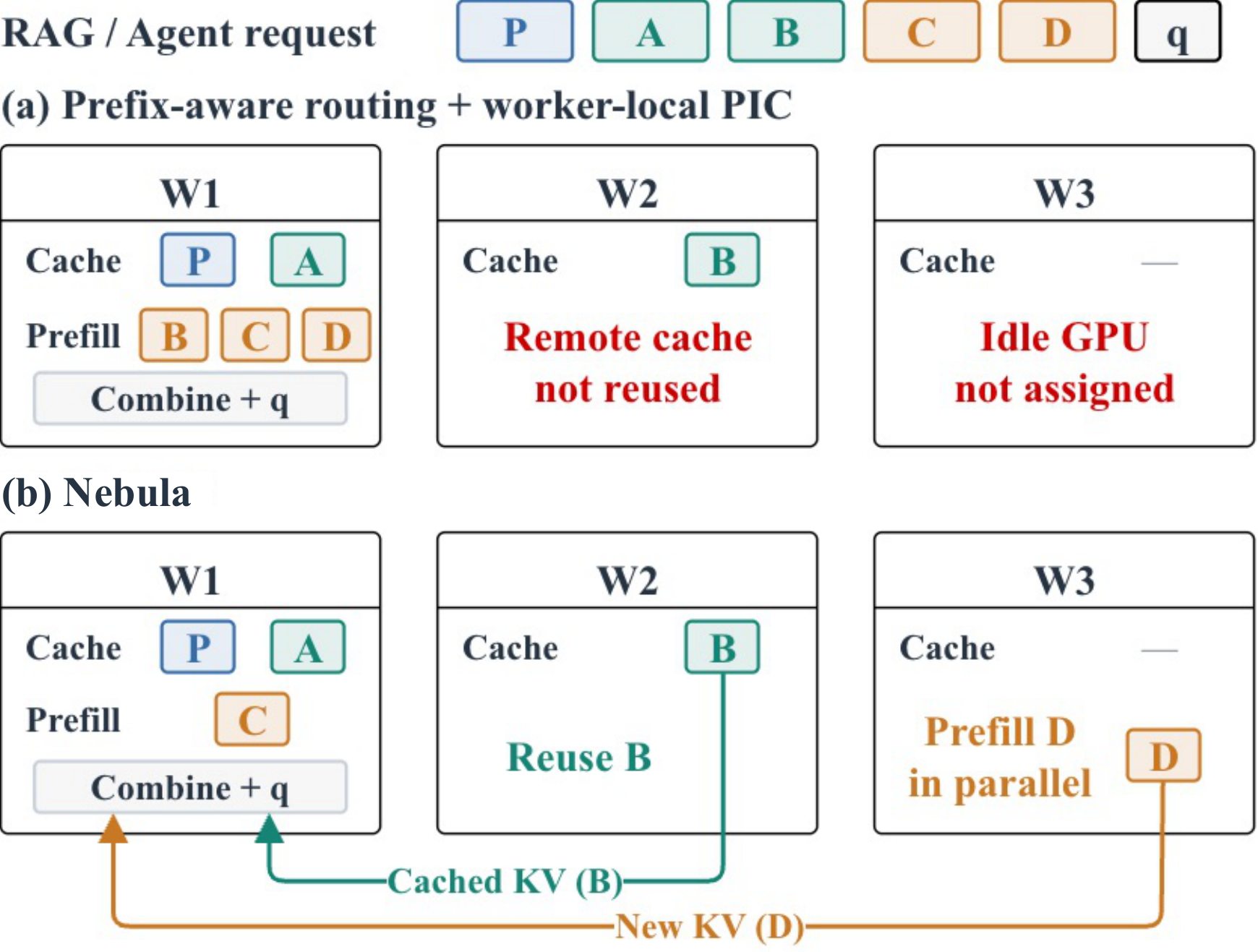}
    \caption{Distributed PIC for a single request. Both plans start with the same cache placement and worker loads, with PIC segments A and B cached on different workers. (a) Worker-local PIC reuses A but leaves remote B and an idle worker unused. (b) \sys reuses B across workers and computes C and D on separate workers while preserving prefix reuse. W1 combines the KV cache and processes query segment $q$.}
    \label{fig:intro-motivation}
\end{figure}

Large language model (LLM) serving is expanding beyond single-turn chat to support retrieval-augmented generation (RAG) and long-horizon agents.
RAG inserts retrieved documents into prompt templates~\cite{choi2018quac,feng2020doc2dial,liu2024chatqa,wang2025ragpulse}, while long-horizon agents extend their context with tool outputs, memory and skill files~\cite{yao2025tau,qian2024chatdev,jimenez2024swe}.
Prefilling these growing contexts requires substantial GPU computation and increases time-to-first-token (TTFT).
Reusing KV caches to avoid repeated prefill computation is therefore important for lowering serving costs and improving responsiveness.

Distributed inference systems extend KV caching across the cluster.
Mooncake~\cite{qin2025mooncake}, LMCache~\cite{liu2025lmcache}, and MemServe~\cite{hu2024memserve} support cross-node indexing, storage, and transfer, allowing requests to reuse KV generated on other nodes and benefit from aggregate cache capacity.
Yet their distributed reuse still relies on strict prefix caching (PC): only content that matches continuously from the beginning of a request can be reused.
PC therefore cannot reuse a previously seen document or tool result independently when its preceding context changes.

In parallel, position-independent caching (PIC)~\cite{hu2025epic, wang2025mepic, yao2025cacheblend, liu2026cacheslide, wang2026prophetkv, ma2025blockattention, lu2025turborag} expands the scope of reuse.
Methods such as CacheBlend~\cite{yao2025cacheblend} and EPIC~\cite{hu2025epic} cache semantically independent segments and \emph{combine} their KV under a different prefix with correction or recomputation as needed.
This matches the recombination of retrieved documents in RAG and the reuse of shared content in different agent contexts.
Across six real-world RAG and agent workloads, our reuse breakdown shows that PIC can recover approximately 72.1\% of the tokens missed by PC on average (\S\ref{sec:pic-opportunity}).

However, simply combining existing PIC algorithms with distributed inference systems does not realize the full capabilities of distributed PIC.
Routers still match caches by prefix and cannot locate cached PIC segments across nodes; schedulers still treat each request as monolithic and cannot parallelize prefill of its cold segments across workers.
Figure~\ref{fig:intro-motivation} illustrates the resulting missed opportunities for remote cache reuse and parallel prefill.

We present \sys, the first distributed LLM serving framework to treat PIC segments as first-class citizens.
\sys enables remote PIC reuse and parallel prefill of uncached segments across workers while preserving PC semantics.
Through a unified backend interface, \sys supports multiple PIC algorithms and decouples algorithm-specific KV processing from cluster-wide resource management.
Building on this framework, we further introduce three designs, each addressing a distinct challenge posed by distributed PIC.

\phm{C1: Keeping the cluster view timely and truthful through optimistic, event-driven view synchronization.}
Distributed PIC can break a single request into hundreds of scheduling units whose states change asynchronously across workers, while routing decisions and actual execution are separated in time.
Inferring cache state solely from scheduling decisions may report invalid cache hits, while relying solely on execution feedback may lead to redundant computation.
\sys combines these complementary sources: the router immediately records expected cache locations and compute load upon dispatch, and workers confirm or correct these records through cache residency, eviction, and task completion events.
This allows \sys to maintain an accurate and timely cluster view, providing a common basis for cache management and prefill scheduling (\S\ref{sec:cluster-view}).

\phm{C2: Balancing PC and PIC allocation through shared, self-tuning cache management.}
Workloads differ in their PC and PIC reuse patterns: RAG documents recur under different prefixes, while successive agent interactions extend the same prefix.
Uncontrolled competition between PC and PIC for shared cache space can cause cache pollution, while fixed quotas can strand cache capacity.
\sys dynamically adjusts the target PC/PIC capacity ratio using cluster-wide ghost-hit feedback, which captures reuse demand after eviction.
Capacity allocation thus adapts to the actual reuse patterns observed across the cluster (\S\ref{sec:cache-management}).

\phm{C3: Balancing compute and transfer through elastic, locality-aware prefill scheduling.}
A distributed PIC request must both fetch cached KV and compute uncached segments.
Placing computation on cache-affine nodes may cause load imbalance, while placing it on idle nodes may increase network traffic.
Optimizing cache locality or load balance alone does not necessarily minimize the prefill completion time of a PIC request.
\sys incorporates queueing, computation, and KV transfer costs into a unified prefill completion time estimate and uses a greedy search to select the best candidate execution plan.
The prefill plans therefore dynamically adjust parallelism and compute placement as load and cache distribution change, accounting for both available compute resources and data locality (\S\ref{sec:scheduling}).

We implement \sys on SGLang~\cite{zheng2023efficiently} and Dynamo~\cite{dynamo_router}, with support for PIC methods including CacheBlend~\cite{yao2025cacheblend}, A$^3$~\cite{zhou2025a3}, EPIC~\cite{hu2025epic} and HYPIC~\cite{liu2026hypic}.
We evaluate \sys with four diverse models on four real-world RAG and agentic workloads.
\sys reduces median TTFT by an average of 61.4\% compared with the strongest existing baseline and recovers an average of 76.1\% of the tokens missed by prefix caching.
\sys's source code will be released at https://github.com/redai-studio/Nebula.

    \section{Background}
\label{sec:background}

\subsection{Prefix Caching at Scale}
\label{sec:prefix-caching}

\phm{Prefix Caching.}
To reduce prefill computation and TTFT for long contexts, mainstream LLM serving systems, such as vLLM~\cite{kwon2023efficient} and SGLang~\cite{zheng2023efficiently}, reuse KV cache across requests with prefix caching (PC).
Under causal attention, each token's KV state depends on that token and its preceding context.
PC therefore requires strict prefix matching: when a new request shares its first $n$ tokens with an earlier request, it can reuse the corresponding KV cache directly with exact numerical equivalence.
Once a prefix mismatch occurs, the KV cache from that point onward becomes immediately unusable.

\phm{Distributed Caching Systems.}
In a serving cluster with multiple LLM instances, reusable KV cache may reside on different nodes and storage tiers, requiring efficient lookup and access during request execution.
Mooncake~\cite{qin2025mooncake}, LMCache~\cite{liu2025lmcache}, and MemServe~\cite{hu2024memserve} support cluster-wide reuse through cache indexing, tiered storage, and KV transfers across nodes, with Mooncake and MemServe further considering cache location and compute load during request scheduling.
TokenLake~\cite{wu2025tokenlake} extends prefix cache management to the segment level, while segment reuse still requires prefix matching.

\subsection{Efficient Position-Independent Caching}
\label{sec:position-independent-caching}

\phm{Position-Independent Caching.}
Position-independent caching (PIC) relaxes strict prefix matching and allows semantically independent segments to be reused at arbitrary positions.
The cached KV still depends on the segment's original position and preceding context, so direct concatenation introduces numerical deviations.
PIC methods typically use recomputation to reduce these deviations: CacheBlend~\cite{yao2025cacheblend} recomputes tokens with the largest KV deviations; A$^3$~\cite{zhou2025a3} identifies key tokens for recomputation from attention distributions; EPIC~\cite{hu2025epic} uses attention sink patterns to statically select the first $k$ tokens of each chunk for recomputation.
HYPIC~\cite{liu2026hypic} further enables PIC for linear attention through transition matrices.

\phm{System Gap.}
However, existing distributed caching systems are built on prefix caching, and none provides complete distributed support for PIC.
CacheBlend implements PIC on top of LMCache but does not integrate it into distributed cache management.
MEPIC~\cite{wang2025mepic} and MiniPIC~\cite{ordonez2026minipic} explore practical PIC implementations but do not consider distributed support.
HYPIC is the closest work to ours, but it uses distributed computation only to accelerate cold prefill and does not address the new challenges that PIC introduces for distributed caching systems (\S\ref{sec:challenges}).

    \section{Analysis}
\label{sec:motivation}

As discussed in \S\ref{sec:background}, PIC can reuse non-prefix content that PC cannot, yet no existing system fully supports distributed PIC.
Before diving into \sys design (\S\ref{sec:solution}), we first examine two questions:
(i) How much additional reuse can PIC provide on top of prefix caching for RAG and agent workloads (\S\ref{sec:pic-opportunity})?
(ii) Is deploying single-node PIC within existing distributed prefix caching systems sufficient to exploit the cluster's cache and compute resources (\S\ref{sec:distributed-pic})?

\subsection{The Opportunity for PIC}
\label{sec:pic-opportunity}

\begin{table}[t]
    \centering
    \footnotesize
    \setlength{\tabcolsep}{2.5pt}
    \begin{tabular}{@{}llrrrr@{}}
        \toprule
        Category & Workload & PC\% & PIC\% & NPHR\% & PMRR\% \\
        \midrule
        RAG & ChatRAG-doc2dial~\cite{liu2024chatqa,feng2020doc2dial} & 2.63 & 99.68 & 97.06 & \textbf{99.7} \\
        RAG & ChatRAG-QuAC~\cite{liu2024chatqa,choi2018quac} & 3.15 & 87.50 & 84.36 & \textbf{87.1} \\
        RAG & RAGPulse~\cite{wang2025ragpulse} & 31.53 & 72.01 & 40.48 & \textbf{59.1} \\
        \midrule
        Agent & nebius-SWE-agent~\cite{badertdinov2024scaling} & 95.26 & 97.49 & 2.23 & \textbf{47.0} \\
        Agent & $\tau$-bench-Retail~\cite{yao2025tau} & 92.62 & 96.69 & 4.07 & \textbf{55.1} \\
        Agent & ChatDev~\cite{qian2024chatdev} & 77.59 & 96.57 & 18.98 & \textbf{84.7} \\
        \bottomrule
    \end{tabular}
    \caption{Theoretical cache hit analysis of PC and PIC on RAG and agent workloads.}
    \label{tab:pic-opportunity}
\end{table}

\phm{PIC Benefit on Prefix Caching.}
Although PC yields low hit rates on RAG workloads, existing production systems already achieve cache hit rates above 90\% on agent workloads.
A meaningful assessment of PIC’s incremental benefits for RAG and agent workloads must therefore account for what existing systems already achieve.
Let $h_{\text{PC}}$ and $h_{\text{PIC}}$ denote the fractions of a request's input tokens reusable through prefix caching and position-independent caching, respectively.
In this theoretical matching analysis, a prefix is a special case of a segment at the beginning of the sequence, so $h_{\text{PIC}} \ge h_{\text{PC}}$.
We use two metrics to quantify PIC's additional benefit over prefix caching:
\begin{itemize}
    \item Non-prefix hit rate (NPHR) $= h_{\text{PIC}} - h_{\text{PC}}$: the fraction of input tokens that miss prefix matching but hit position-independent matching, or the absolute increase in hit rate provided by PIC.
    \item Prefix-miss recovery rate (PMRR) $= \text{NPHR}\,/\,(1 - h_{\text{PC}})$: the fraction of prefix-missed tokens that PIC can reuse, directly reflecting how much computation can be further reduced after prefix caching has been fully exploited.
\end{itemize}
We select six public workloads spanning RAG and agent scenarios.
For RAG, ChatRAG-doc2dial~\cite{liu2024chatqa,feng2020doc2dial} and ChatRAG-QuAC~\cite{liu2024chatqa,choi2018quac} are conversational RAG workloads, and RAGPulse~\cite{wang2025ragpulse} is a public RAG serving trace.
For agents, nebius-SWE-agent~\cite{badertdinov2024scaling} is a coding agent trace, $\tau$-bench-Retail~\cite{yao2025tau} is a tool-use agent trace, and ChatDev~\cite{qian2024chatdev} is a multi-agent software collaboration trace.

As shown in Table~\ref{tab:pic-opportunity}, RAG and agent workloads exhibit markedly different reuse patterns.
RAG requests concatenate independently retrieved texts after a system prompt, often leaving repeated content outside the shared prefix.
ChatRAG-doc2dial and ChatRAG-QuAC have prefix hit rates of only about 3\%, yet their NPHR reaches 97\% and 84\%, respectively, with most hits coming from non-prefix reuse.
Even RAGPulse, with a higher prefix hit rate of 31.5\%, has a PMRR of 59.1\%.

Agent requests grow over successive interactions, giving tool-use and coding workloads high prefix hit rates: 92.6\% for $\tau$-bench-Retail and 95.3\% for nebius-SWE-agent, with NPHR of only 2\% to 4\%.
Yet the dominance of prefix reuse does not mean that PIC has no value: for the less than 10\% of input tokens that miss the prefix cache, PMRR is still 55.1\% and 47.0\% in these two workloads, respectively.
A role-level breakdown further shows that environment observations account for 85.4\% of the non-prefix reuse in nebius-SWE-agent, while tool-returned database records account for 89.5\% in $\tau$-bench-Retail.
\textbf{This means that even with such high prefix hit rates, nearly half of the remaining computation can still be saved by PIC.}
In the multi-agent workload ChatDev, content shared across agents appears at different positions, also creating considerable PIC reuse opportunities with a PMRR of 84.7\%.

Despite their different reuse patterns, both RAG and agent workloads can save a substantial fraction of prefill computation by adding PIC to existing PC.

\subsection{From PIC to Distributed PIC}
\label{sec:distributed-pic}

\begin{figure}
    \centering
    \subfloat[Cache hit ratio.]{
        \label{fig:moti-chr}
        \centering
        \includegraphics[width=\linewidth]{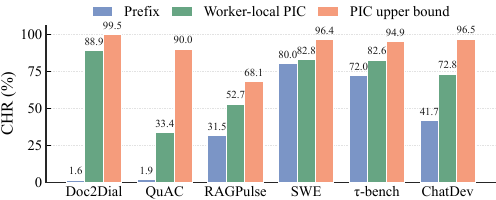}
    }
    \hfill
    \subfloat[Tail TTFT of worker-local PIC.]{
        \label{fig:moti-ttft}
        \centering
        \includegraphics[width=\linewidth]{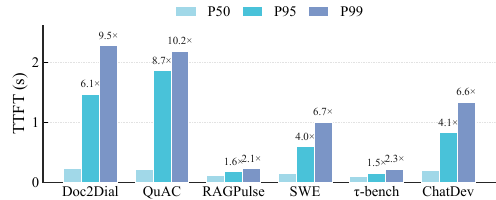}
    }
    \caption{Limitations of deploying PIC-enabled workers behind a prefix-cache-aware router. (a) Worker-local PIC improves cache hit ratio (CHR) over PC but remains below the ideal PIC upper bound. (b) Worker-local PIC exhibits substantial tail TTFT across the same workloads.}
    \vspace{-2ex}
    \label{fig:moti-gap}
\end{figure}

\phm{The Gap between PIC and Distributed PIC.}
To assess the gap between the existing system and the ideal state, we first construct a direct-composition strawman by deploying PIC-enabled workers behind Dynamo, an existing prefix-cache-aware router.
On the one hand, this setup lacks \textbf{remote cache reuse}: it cannot locate cached PIC segments on other nodes, so it may recompute content that could instead be fetched over the network for reuse.
On the other hand, this setup lacks \textbf{parallel prefill}: it cannot offload independently computable PIC segments to idle nodes, so it may leave available cluster compute resources unused and suffer from high TTFT.

We deploy the worker-local PIC strawman in a cluster with seven prefill workers and one decode worker, and replay the six workloads at 8 requests/s in their original order from a cold cache.
Figure~\ref{fig:moti-gap} (a) shows that worker-local PIC improves CHR over PC, yet remains 10.6--56.7 percentage points below the ideal PIC upper bound across the six workloads.
The gap is largest on ChatRAG-QuAC, where worker-local PIC achieves only 33.4\% CHR against an upper bound of 90.0\%.
Figure~\ref{fig:moti-gap} (b) further shows substantial tail TTFT: P99 is 2.1--10.2$\times$ P50, reaching 9.5$\times$ on ChatRAG-doc2dial and 10.2$\times$ on ChatRAG-QuAC.
Because each request remains bound to one worker, the strawman cannot use idle workers to accelerate the prefill of its uncached PIC segments.

These observations motivate a distributed PIC system that can fully exploit cluster-wide cache and compute resources.

    \section{\sys Overview}
\label{sec:solution}

We present \sys, the first distributed LLM serving system to treat PIC segments as first-class citizens.
Distributed PIC is more than introducing a new cache object: it entails joint, system-level optimization across view synchronization, cache management, and prefill scheduling.
This section first presents the design principles (\S\ref{sec:design-principles}) and system architecture (\S\ref{sec:architecture}), then outlines three new challenges posed by distributed PIC (\S\ref{sec:challenges}).
We then detail the three corresponding solutions: Optimistic, Event-Driven View Synchronization (\S\ref{sec:cluster-view}), Shared, Self-Tuning Cache Management (\S\ref{sec:cache-management}), and Elastic, Locality-Aware Prefill Scheduling (\S\ref{sec:scheduling}).

\subsection{Design Principles}
\label{sec:design-principles}

\phm{Treat PIC segments as first-class citizens.}
Existing PIC works identify and reuse PIC segments only within individual workers, whereas existing distributed caching systems still manage prefix caches across nodes.
Simply composing the two neither enables cross-node reuse of cached PIC segments nor allows independently computable PIC segments to be offloaded to idle workers (\S\ref{sec:distributed-pic}).
A PIC segment must therefore be more than a piece of KV state internal to the inference engine.
It must become a first-class distributed cache object and the common unit of operation for view synchronization, cache management, and prefill scheduling.
Each PIC segment can then be independently indexed, placed, transferred, reused, or computed across the cluster, extending PIC from a worker-local optimization to a cluster-wide capability.

\phm{Complement prefix caching rather than replace it.}
The relative benefits of PC and PIC vary across workloads (\S\ref{sec:pic-opportunity}), while rebuilding prefix cache with recomputation-based PIC still incurs overhead~\cite{yao2025cacheblend, zhou2025a3, hu2025epic, liu2026hypic}.
A distributed PIC framework should therefore preserve existing prefix-caching semantics and add PIC as a complementary capability.
The two cache types should share the same storage and transfer infrastructure while retaining their distinct matching semantics.

\phm{Make PIC algorithms pluggable.}
Different PIC algorithms handle segment reuse and recomputation in different ways (\S\ref{sec:position-independent-caching}).
A distributed PIC framework should therefore not be tied to a single algorithm.
Instead, it should integrate existing PIC algorithms as pluggable backends through a unified interface that decouples cluster-wide cache management and prefill scheduling from algorithm-specific KV processing, enabling single-node PIC algorithms to operate in distributed settings.

\sys leaves KV reuse and recomputation policies, including their accuracy tradeoffs, to the PIC backend.
It adapts different PIC backends for distributed execution and focuses on cluster-wide resource management.

\subsection{System Architecture}
\label{sec:architecture}

\begin{figure}[t]
    \centering
    \includegraphics[width=\linewidth]{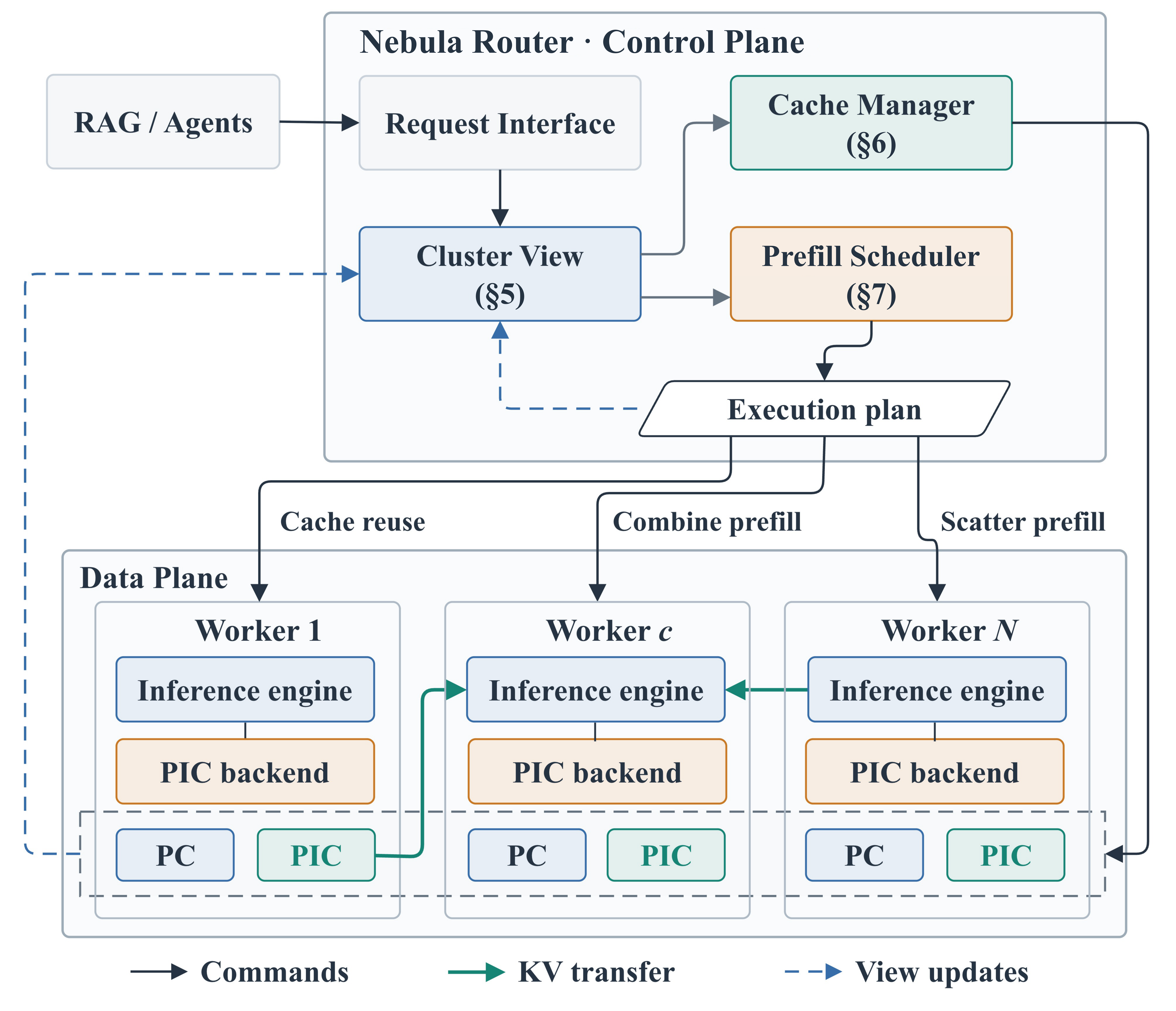}
    \caption{\sys system architecture.}
    \label{fig:system-architecture}
\end{figure}

\phm{Framework components.}
As shown in Figure~\ref{fig:system-architecture}, \sys consists of a control plane and a data plane.
The control plane makes cluster-wide cache and compute decisions, and the data plane executes them.

The control plane is implemented by a cluster-level router and contains three components:
\begin{componentlist}
    \item \textbf{Cluster View} maintains cluster-wide cache state and load state, providing a unified view for cache management and prefill scheduling (\S\ref{sec:cluster-view}).
    \item \textbf{Cache Manager} uses this view to allocate shared cache capacity between PC and PIC (\S\ref{sec:cache-management}).
    \item \textbf{Prefill Scheduler} uses the same view to construct execution plans (\S\ref{sec:scheduling}).
\end{componentlist}
The data plane comprises two PIC-enhanced components:
\begin{componentlist}
    \item \textbf{Cache Pool} retains existing prefix cache storage and transfer capabilities and extends them to PIC segments.
    \item \textbf{Inference Workers} retain their existing model execution paths and invoke specific PIC algorithms through a unified backend interface.
\end{componentlist}

\begin{figure}[t]
    \centering
    \begin{tabular}{@{}c@{}}
    \tikz \node[draw, rounded corners=2pt, align=left,
        text width=.91\columnwidth, inner sep=3pt, font=\scriptsize] {
        \textbf{RAG application input}\\
        \emph{System:} You are a customer-support assistant.\\
        \emph{User:} \texttt{<pic>}A damaged product may be returned within 30 days.\texttt{</pic>}\\
        \hspace*{2.9em}\texttt{<pic>}Refunds use the original payment method.\texttt{</pic>}\\
        \hspace*{2.9em}Question: How long do I have to return a damaged product?
    };\\[0.6ex]
    \tikz \node[draw, rounded corners=2pt, align=left,
        text width=.91\columnwidth, inner sep=3pt, font=\scriptsize] {
        \textbf{Agent application input}\\
        \emph{System, user, assistant:} coding instructions and conversation history\\
        \emph{Tool:} \texttt{<pic>}[parser.py] \texttt{def parse\_config(path): ...}\texttt{</pic>}\\
        \emph{Tool:} \texttt{<pic>}[SKILL.md] Normalize keys before validation.\texttt{</pic>}\\
        \emph{User:} Update the parser to follow these guidelines.
    };
    \end{tabular}
    \caption{Application-defined text segmentation.}
    \label{fig:pic-marking}
\end{figure}

\phm{Application Interface and PIC Segments.}
\sys lets applications define where PIC is used.
The application partitions each request into text segments: content enclosed by the PIC tags \texttt{<pic>...</pic>} forms an application-defined PIC segment, while each maximal unmarked span forms a regular segment.
The interface removes these tags before tokenization.
Unlike PIC segments, regular segments cannot be independently prefilled, cached, transferred, or reused across the cluster.
Instead, they are reused through the prefix cache or recomputed in the request context on a prefix-cache miss.

Figure~\ref{fig:pic-marking} shows example segmentations for RAG and agent requests.
RAG applications mark retrieved documents as PIC segments in their request templates, while agent applications mark tool outputs, environment observations, and cross-agent messages as PIC segments.
The remaining unmarked content forms regular segments.
System prompts typically form stable request prefixes and can be reused through prefix caching, so they remain unmarked in these examples.

\phm{Request lifecycle.}
When a request arrives, the control plane first performs prefix matching over the complete request, then position-independent matching for PIC segments not covered by the prefix hit.
The Prefill Scheduler uses the matching results and worker load to construct an execution plan.
The plan specifies a worker for each PIC segment, either to supply cached KV or to perform prefill, and a \emph{combine home}, where the required KV is gathered to complete the request's prefill.
We detail plan generation in \S\ref{sec:scheduling}.

\begin{figure}[t]
    \centering
    \includegraphics[width=0.9\linewidth]{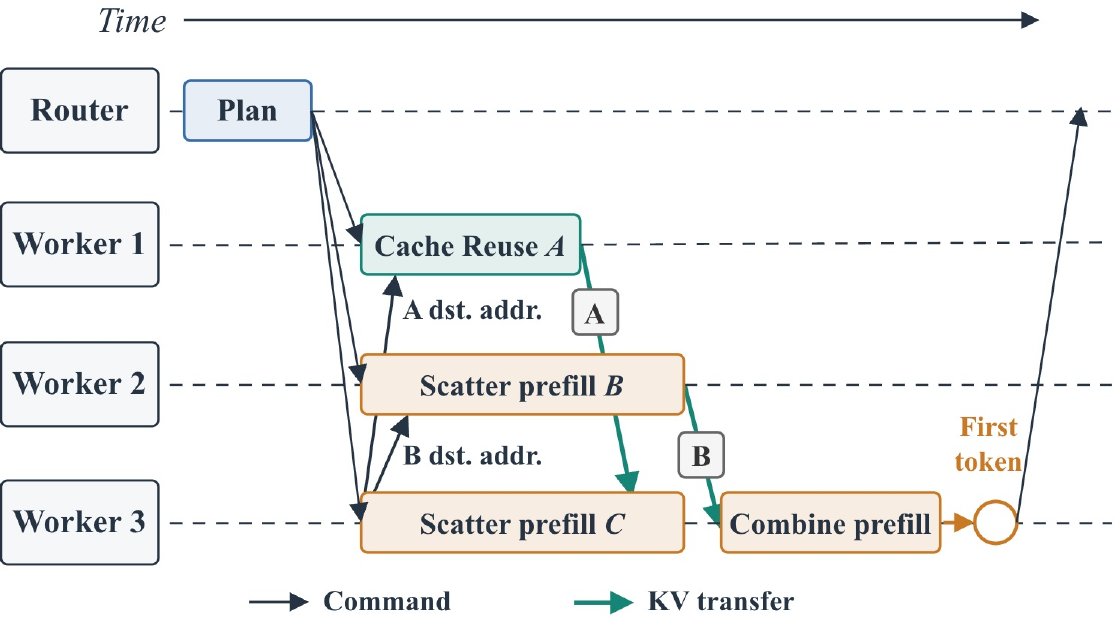}
    \caption{Execution timeline for one PIC request.}
    \label{fig:request-lifecycle}
\end{figure}

Figure~\ref{fig:request-lifecycle} illustrates execution through three types of operations.
\textbf{Scatter prefill} independently computes cold PIC segments on assigned workers and transfers their generated KV to the combine home when needed.
\textbf{Cache reuse} makes matched PC/PIC KV available at the combine home, transferring remotely cached KV when needed.
\textbf{Combine prefill} invokes the PIC backend at the combine home to reuse PIC segments, compute the remaining content, and perform necessary recomputation.
Scatter workers can begin computation once they receive their tasks, but transfer generated KV via direct GPU writes only after receiving destination GPU addresses from the combine home.
Scatter prefill, cache reuse, and control message delivery can overlap, while combine prefill begins only after all required KV is available at the combine home.

Once prefill completes, \sys hands off the request and its KV state to a decode worker through the standard PD-disaggregated serving path. \sys leaves the decode path unchanged, focusing on reducing TTFT through prefill optimization.

\subsection{Challenges}
\label{sec:challenges}

The framework above extends PIC from a single worker to the entire cluster.
However, making the system efficient remains non-trivial: distributed PIC introduces three new system challenges in view synchronization, cache management, and prefill scheduling.

\phm{C1: Keeping the Cluster View Timely and Truthful under Fine-Grained Reuse.}
Distributed PIC substantially increases the number of independently scheduled units whose cache and compute states change asynchronously across workers, while routing decisions and actual execution are separated in time.
An inaccurate view can cause invalid cache hits, whereas delayed updates can lead to redundant computation.
The cluster view must therefore be both timely and truthful to support cache management and prefill scheduling (\S\ref{sec:cluster-view}).

\phm{C2: Balancing PC and PIC Allocation under a Shared Budget.}
PC and PIC share finite cache capacity, but their relative reuse demand varies across workloads.
Uncontrolled competition between them can crowd out frequently reused content, while fixed partitions can leave capacity idle for one type even as the other runs out of space.
Cache management must therefore dynamically coordinate PC and PIC allocation as cluster-wide reuse demand changes (\S\ref{sec:cache-management}).

\phm{C3: Balancing Compute and Transfer under Shifting Load.}
Remote cache reuse incurs network transfer delays, while parallel prefill must wait for the slowest computation to finish.
Under changing load, placing computation on cache-affine nodes may increase queueing delays, while placing it on less-loaded nodes may increase KV transfer costs.
Scheduling must therefore balance compute and transfer when jointly selecting compute placement and parallelism (\S\ref{sec:scheduling}).

    \section{Optimistic, Event-Driven View Synchronization}
\label{sec:cluster-view}

\subsection{Problem Analysis}
\label{sec:cluster-view-challenge}

\begin{table}[t]
    \centering
    \footnotesize
    \begin{tabular}{@{}lrrrrrr@{}}
        \toprule
        & \multicolumn{3}{c}{Segments/request}
        & \multicolumn{3}{c}{Length (tokens)} \\
        \cmidrule(lr){2-4}\cmidrule(lr){5-7}
        Workload & P50 & P95 & P99 & P50 & P95 & P99 \\
        \midrule
        ChatRAG-doc2dial & 106 & 112 & 113 & 334 & 670 & 1,190 \\
        ChatRAG-QuAC & 35 & 127 & 203 & 536 & 773 & 939 \\
        RAGPulse & 11 & 13 & 15 & 319 & 521 & 521 \\
        \midrule
        nebius-SWE-agent & 31 & 253 & 413 & 69 & 1,032 & 1,431 \\
        $\tau$-bench-Retail & 14 & 34 & 44 & 93 & 1,240 & 1,240 \\
        ChatDev & 22 & 65 & 73 & 818 & 2,253 & 2,780 \\
        \bottomrule
    \end{tabular}
    \caption{Distributions of segment counts and lengths across RAG and agent workloads.}
    \label{tab:segment-distribution}
    \vspace{-2ex}
\end{table}

\begin{figure}[t]
    \centering
    \includegraphics[width=0.95\columnwidth]{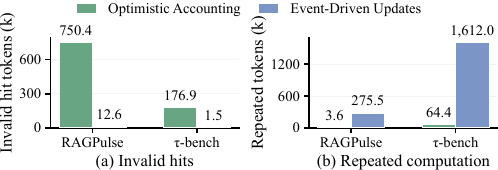}
    \caption{Limitations of Optimistic Accounting and Event-Driven Updates used in isolation: (a) invalid hit tokens and (b) repeated computation tokens on RAGPulse and $\tau$-bench-Retail.}
    \label{fig:view-motivation}
    \vspace{-2ex}
\end{figure}

Distributed PIC divides each request into multiple PIC segments that can be independently placed, transferred, and computed.
This shift from request-level to segment-level scheduling increases the number of scheduling units by $11$--$106\times$ at the median and up to $413\times$ at P99 (Table~\ref{tab:segment-distribution}).
Dispatched segment computations may still be queued or running when subsequent requests arrive, while cache insertions and evictions occur asynchronously across workers.

We construct two strawmen for distributed-PIC cache-directory maintenance based on existing view-synchronization mechanisms:
(i) \textbf{Optimistic Accounting} infers PIC cache state from scheduling decisions, worker capacity, and an LRU model, following the SGLang router~\cite{sglang_router} and vLLM router~\cite{vllm_router}.
(ii) \textbf{Event-Driven Updates} maintains the PIC cache index using worker-reported cache events, following Dynamo~\cite{dynamo_router} and llm-d~\cite{llmd_kv_indexer}.
However, the optimistic strawman exposes assigned work immediately but can report invalid cache hits when its predicted PIC-segment locations drift from actual residency.
The event-driven strawman does not assume cache availability before worker confirmation, which can lead to redundant prefill of the same segment across requests.

We compare the two variants using Qwen3.5-35B-A3B on a 7P1D cluster, replaying the RAGPulse and $\tau$-bench-Retail traces at 40 QPS.
We measure \emph{invalid hit tokens}, where expected holders have neither usable KV nor earlier in-flight work, and \emph{repeated computation tokens}, where directory misses trigger redundant segment prefill despite earlier in-flight work.
Figure~\ref{fig:view-motivation} shows that Optimistic Accounting incurs $59.4\times$ and $114.3\times$ as many invalid hit tokens as Event-Driven Updates on the two workloads, respectively.
Conversely, Event-Driven Updates incurs $77.3\times$ and $25.0\times$ as many repeated computation tokens.

Therefore, under fine-grained PIC reuse, view synchronization cannot rely on router-side speculation or worker-side confirmation alone.
It must address both view drift and delayed updates, providing a basis that is both timely and truthful for cache management and prefill scheduling.

\subsection{Our Solution}
\label{sec:cluster-view-design}

The key insight of \sys is to combine the router's immediate scheduling decisions with workers' execution feedback.

\begin{figure}[t]
    \centering
    \includegraphics[width=0.95\linewidth]{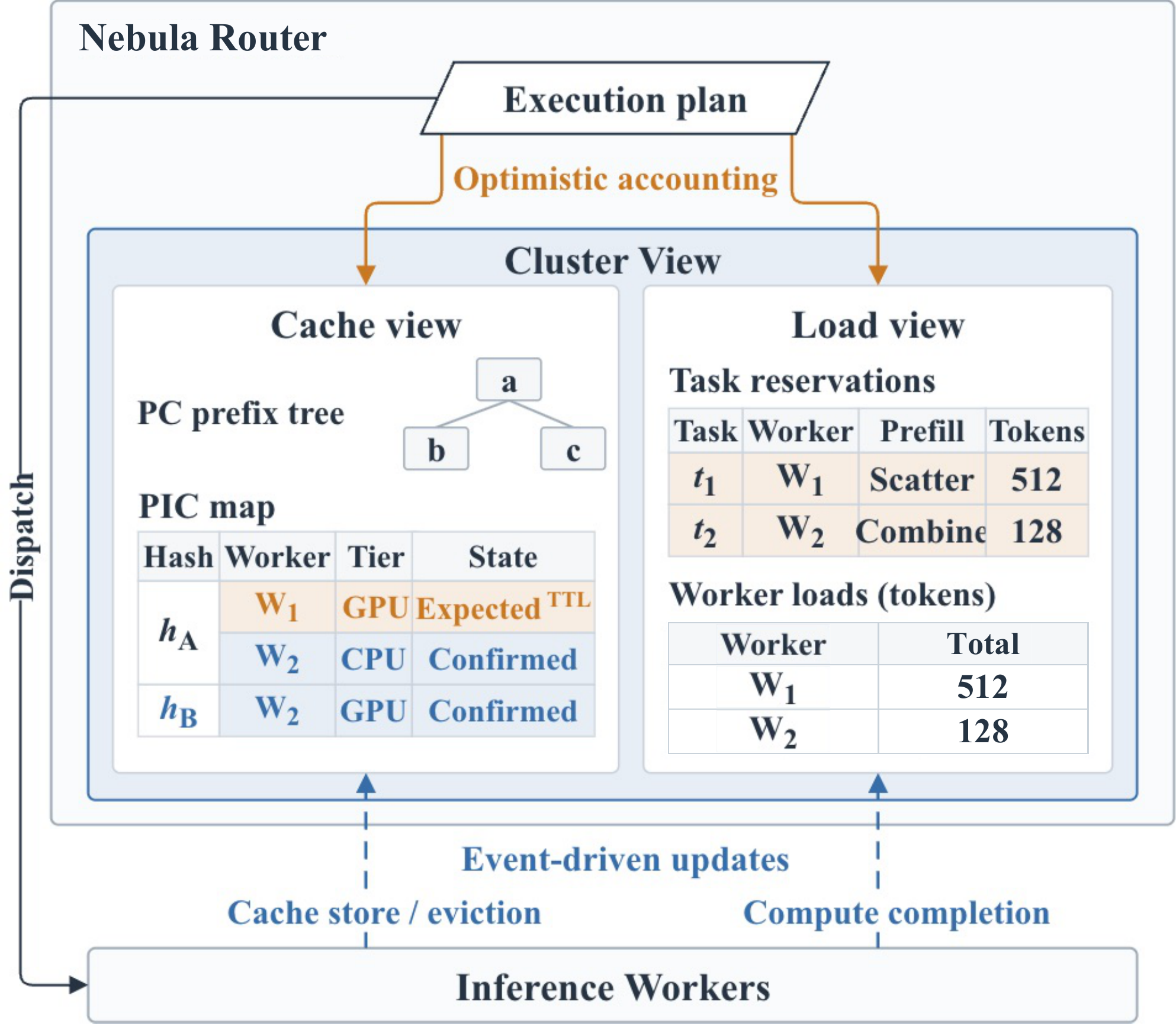}
    \caption{View synchronization in \sys.}
    \label{fig:view-sync}
    \vspace{-2ex}
\end{figure}

\phm{View Organization.}
The cluster view of \sys consists of a cache view and a load view (Figure~\ref{fig:view-sync}).
The cache view retains the existing PC prefix index and organizes the PIC map by content hash.
For each PIC segment, it records its worker locations, storage tiers, and whether each record has been confirmed by a worker-side event.
The load view records each task's target worker and reserved compute work and aggregates these reservations into each worker's outstanding compute load.

\phm{Optimistic Accounting and Event-Driven Updates.}
When dispatching an execution plan, the router immediately records expected PIC-segment locations in the cache directory and reserves compute load for scatter prefill and combine prefill tasks.
Subsequent requests can thus identify and await matching in-flight results on the assigned workers, avoiding redundant prefill.
During execution, workers send incremental events to the router: cache-store events confirm the corresponding locations, cache-eviction events remove invalid locations, and compute-completion reports release the corresponding tasks' load reservations.

\sys assigns each unconfirmed cache-location entry a TTL and removes it upon expiration.
When a cache-store event arrives, the corresponding location becomes confirmed and is no longer subject to the TTL.
Late cache-store events can still re-register the corresponding locations.

Unlike unconfirmed cache-location entries, load reservations do not expire on timeout, because their tasks may still be queued or running.
The router releases a reservation only after confirming that the task has completed or will not execute, and deduplicates completion reports so that each reservation is released at most once.

    \section{Shared, Self-Tuning Cache Management}
\label{sec:cache-management}

\subsection{Problem Analysis}
\label{sec:cache-management-challenge}

\begin{figure}[t]
    \centering
    \includegraphics[width=0.95\columnwidth]{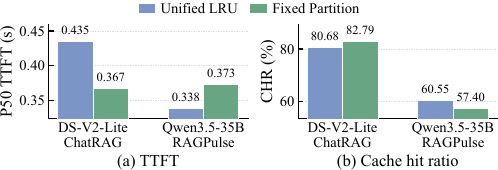}
    \caption{Cache allocation trade-off: (a) median TTFT and (b) token-weighted cache hit ratio.}
    \label{fig:cache-management-challenge}
    \vspace{-2ex}
\end{figure}

Recompute-based PIC methods may store different KV representations of the same segment in PC and PIC, creating competing demands for finite cache capacity.
The relative demand for the two cache types depends on workload reuse patterns (\S\ref{sec:pic-opportunity})---
doc2dial and QuAC reuse RAG documents under different prefixes, whereas nebius-SWE-agent and $\tau$-bench-Retail repeatedly extend the same prefix; RAGPulse and ChatDev exhibit both patterns.

We construct two strawmen for PC--PIC cache allocation based on existing capacity-management policies:
(i) \textbf{Unified LRU}, as used by MiniPIC~\cite{ordonez2026minipic}, manages PC and PIC in a shared cache pool on each worker without cache-type quotas and selects eviction victims by access recency.
(ii) \textbf{Fixed Partition}, following HYPIC's separate capacity budgets and per-pool LRU~\cite{liu2026hypic}, divides capacity between PC and PIC at a preset ratio, with no borrowing between the two pools.
However, Unified LRU allocates shared capacity solely through recency-based eviction, which may allow one cache type to crowd out frequently reused content from the other.
Fixed Partition limits this interference but cannot adapt to changing demand, potentially leaving capacity idle in one pool while the other repeatedly evicts entries for lack of space.

We compare the two policies on a 7P1D cluster using DeepSeek-V2-Lite with ChatRAG-QuAC at 16 QPS and Qwen3.5-35B-A3B with RAGPulse at 40 QPS.
Figure~\ref{fig:cache-management-challenge} shows that Fixed Partition achieves a higher cache hit ratio on ChatRAG (82.79\% versus 80.68\%), while Unified LRU wins on RAGPulse (60.55\% versus 57.40\%).
Neither policy dominates across these two configurations.

Therefore, under a shared cache budget, the Cache Manager cannot rely solely on a unified eviction policy or fixed quotas to allocate capacity.
Instead, it must self-tune the PC and PIC capacity shares as cluster-wide reuse demand changes, improving cache reuse under finite capacity.

\subsection{Our Solution}
\label{sec:cache-management-design}

The key insight of \sys is to use cluster-wide reuse feedback to dynamically adjust the shares of PC and PIC.

\begin{figure}[t]
    \centering
    \includegraphics[width=0.95\linewidth]{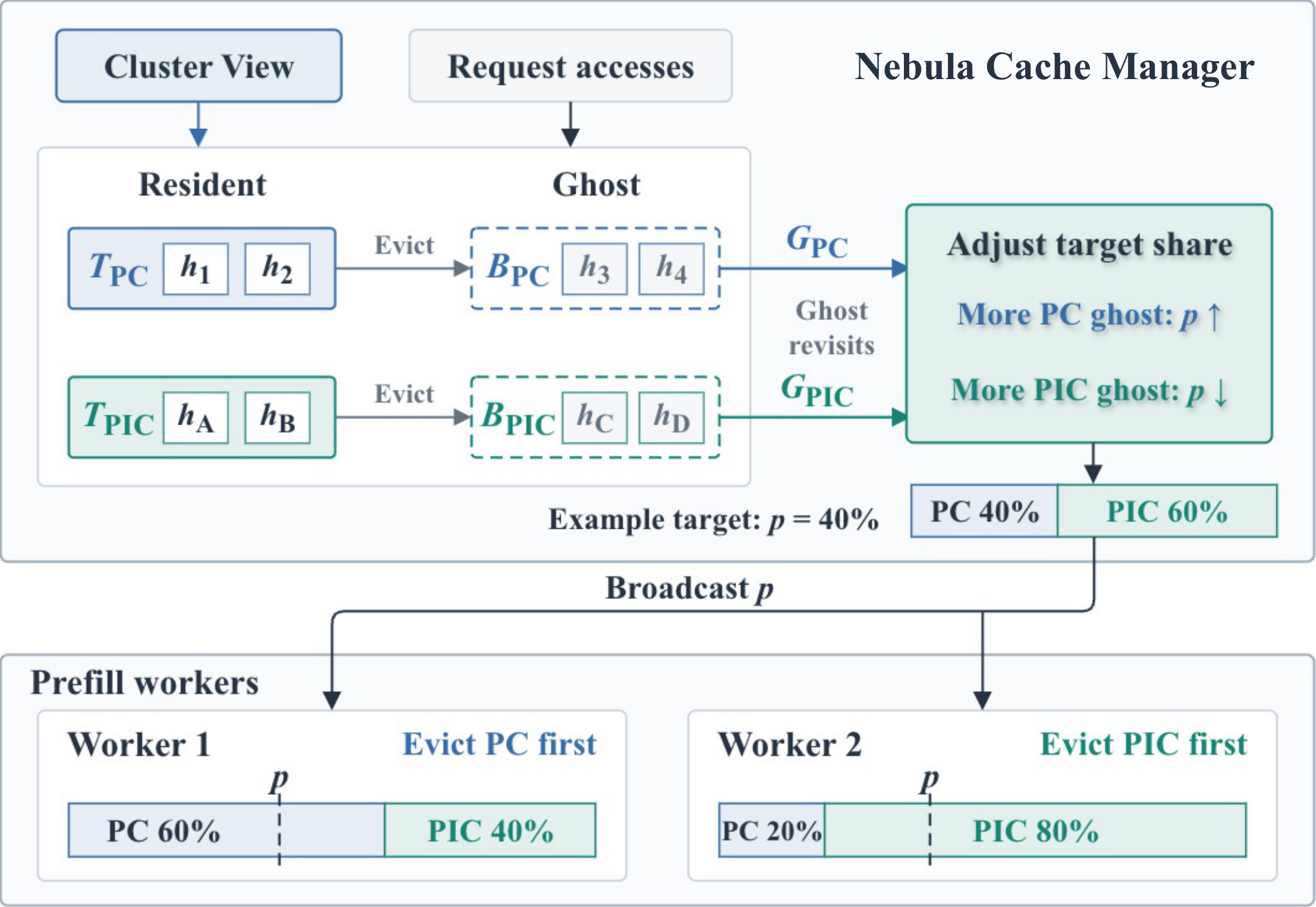}
    \caption{Self-tuning cache management in \sys. }
    \label{fig:cache-management}
    \vspace{-2ex}
\end{figure}

\phm{Cluster-Wide Reuse Feedback.}
Inspired by ARC's adaptation between recency and frequency~\cite{megiddo2003arc}, \sys uses revisits to evicted cache entries to identify which cache type needs more space.
As shown in Figure~\ref{fig:cache-management}, the router maintains four queues: $T_{\mathrm{PC}}$ and $T_{\mathrm{PIC}}$ track PC and PIC entries present in the cluster, while $B_{\mathrm{PC}}$ and $B_{\mathrm{PIC}}$ track previously cached entries no longer present in the cluster.
All four queues store only identifiers, not cached data.

When an entry appears in the cluster view, the router adds it to the corresponding $T$ queue and removes its history from the corresponding $B$ queue.
When the cluster view no longer reports any available holder for an entry, the router moves it from $T$ to $B$.

Reuse feedback follows the prefix-first principle (\S\ref{sec:design-principles}).
The router first matches existing prefix caches in $T_{\mathrm{PC}}$, then checks $B_{\mathrm{PC}}$ to retain revisit feedback for missing prefix caches, even when PIC can still serve the same content.
It then checks $T_{\mathrm{PIC}}$ and $B_{\mathrm{PIC}}$ in order to assess PIC reuse demand.
Each token in a request contributes to at most one feedback category.

\phm{Self-Tuning Capacity Shares.}
Let $G_{\mathrm{PC}}$ and $G_{\mathrm{PIC}}$ denote the token counts covered by ghost revisits for the two cache types.
Let $p\in[0,1]$ be the global target soft share for PC, with $1-p$ for PIC.
The router updates this share over request windows:
\begin{equation}
\label{eq:cache-share-update}
\begin{aligned}
g&=\frac{G_{\mathrm{PC}}-G_{\mathrm{PIC}}}{G_{\mathrm{PC}}+G_{\mathrm{PIC}}},\\
p_{\mathrm{new}}&=\operatorname{clip}_{[0,1]}\left[p+0.01g\right].
\end{aligned}
\end{equation}
More PC ghost tokens increase $p$, while more PIC ghost tokens decrease it.

\sys initializes $p$ to $0.5$ and updates it every $U=16$ requests, changing it by at most one percentage point per update.
After each update, the router asynchronously broadcasts the new allocation ratio to all prefill workers.

\phm{Worker-Side Soft Allocation.}
All prefill workers use the same target ratio.
Let $R_{\mathrm{PC}}$ and $R_{\mathrm{PIC}}$ denote the local cache occupancy of PC and PIC, respectively.
If
\begin{equation}
\label{eq:cache-eviction-priority}
R_{\mathrm{PC}}>p\bigl(R_{\mathrm{PC}}+R_{\mathrm{PIC}}\bigr),
\end{equation}
the worker prioritizes PC eviction; otherwise, it prioritizes PIC eviction.
Updating $p$ changes only the cache-type priority for subsequent reclamation and triggers no immediate eviction.

For tiered storage, \sys maintains the four queues separately for each storage tier and adjusts its PC/PIC target shares using that tier's reuse feedback.
Workers independently apply soft allocation at each tier based on its target shares.

    \section{Elastic, Locality-Aware Prefill Scheduling}
\label{sec:scheduling}

\subsection{Problem Analysis}
\label{sec:scheduling-challenge}

\begin{figure}[t]
    \centering
    \includegraphics[width=0.95\columnwidth]{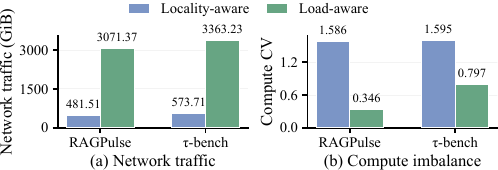}
    \caption{Locality--load trade-off on RAGPulse and $\tau$-bench-Retail. Locality-aware reduces network traffic, while Load-aware lowers the CV of cumulative prefill tokens across workers.}
    \label{fig:scheduling-challenge}
    \vspace{-2ex}
\end{figure}

Distributed PIC allows cold PIC segments to be prefilled in parallel across workers, while PC and PIC may reside on different workers.
Under changing cluster load, scheduling must jointly select each cache's reuse path, the combine home and the placement of cold PIC-segment computation so that the required KV becomes ready at the combine home as early as possible for the subsequent combine prefill.

Building on existing cache-aware and load-balancing approaches, we construct two strawman schedulers for distributed PIC:
(i) \textbf{Locality-aware} selects the combine home with the most local PC/PIC hits, computes cold PIC segments there, and fetches the remaining PIC hits from other workers.
(ii) \textbf{Load-aware} assigns scatter and combine prefill tasks to workers using longest-processing-time-first (LPT) scheduling and transfers both cached and generated KV from remote workers to the combine home.
However, Locality-aware may increase queueing delays by placing computation on a cache-affine but busy worker.
Load-aware may incur higher transfer costs when fetching cache hits or returning newly generated KV.
Having more local cache hits or more balanced computation does not necessarily lead to earlier prefill completion.

We compare the two schemes using Qwen3.5-35B-A3B on a 7P1D cluster, replaying the RAGPulse and $\tau$-bench-Retail traces at 8 QPS.
We measure total network traffic and quantify compute imbalance using the population coefficient of variation (CV) of cumulative prefill tokens across prefill workers.
As Figure~\ref{fig:scheduling-challenge} shows, Load-aware transfers $6.4\times$ and $5.9\times$ as much data as Locality-aware on the two workloads, respectively, but reduces compute CV by $4.6\times$ and $2.0\times$.

Therefore, under changing request load and cache distribution, prefill scheduling cannot prioritize cache locality or load balance alone.
Instead, it must jointly consider compute load and KV transfer costs when selecting placement and parallelism to minimize prefill completion time.

\subsection{Our Solution}
\label{sec:scheduling-design}

\begin{algorithm}[t]
\caption{Greedy Prefill Planning}
\label{alg:prefill-planning}
\small
\begin{algorithmic}[1]
    \Require $\mathcal S$, $\mathcal W$, $\{\mathcal H_w,Q_w\}$, $\{P_i\}$, cache-hit sources, $\widehat T(\cdot)$
    \Ensure Execution plan $\pi_{\mathrm{best}}$
    \State $\mathit{cold}\gets$ sort $\mathcal S\setminus\bigcup_{w\in\mathcal W}\mathcal H_w$ by decreasing $P_i$\label{line:prefill-cold}
    \State $\Pi\gets\varnothing$
    \For{each $c\in\mathcal W$}
        \label{line:prefill-combine}
        \State $a\gets$ empty mapping
        \For{each $i\in\mathcal S$}
            \State $a(i)\gets c$ if $i\in\mathit{cold}$, else its cache source
        \EndFor
        \State $\pi\gets(c,a)$\label{line:prefill-init}
        \For{each $i\in\mathit{cold}$}
            \label{line:prefill-scatter}
            \State \textbf{if} $\mathcal W\setminus\{c\}=\varnothing$ \textbf{then break}
            \State $w\gets\displaystyle\operatorname*{arg\,min}_{v\in\mathcal W\setminus\{c\}}\left[Q_v+P_v(\pi)\right]$
            \State $a'\gets\operatorname{copy}(a)$
            \State $a'(i)\gets w$
            \State $\pi'\gets(c,a')$\label{line:prefill-candidate}
            \If{$\widehat T(\pi')<\widehat T(\pi)$}
                \label{line:prefill-accept}
                \State $a\gets a'$
                \State $\pi\gets\pi'$\label{line:prefill-update}
            \EndIf
        \EndFor
        \State $\Pi\gets\Pi\cup\{\pi\}$\label{line:prefill-best}
    \EndFor
    \State $\pi_{\mathrm{best}}\sim\operatorname{Uniform}\!\left(\operatorname*{arg\,min}_{\pi\in\Pi}\widehat T(\pi)\right)$
    \State \Return $\pi_{\mathrm{best}}$\label{line:prefill-return}
\end{algorithmic}
\end{algorithm}

The key insight of \sys is to jointly account for queueing, computation, and KV transfer when estimating a distributed PIC request's prefill completion time.

\phm{Estimating Prefill Completion Time.}
\sys constructs execution plans in request arrival order (FIFO).
Each plan consists of a combine home $c$ and a mapping $a$ from the request's PIC segments to workers:
\begin{equation}
    \pi=(c,a),\qquad a:\mathcal S\rightarrow\mathcal W,\quad c\in\mathcal W,
\end{equation}
where $\mathcal S$ is the request's PIC segment set and $\mathcal W$ is the cluster's worker set.

To estimate the completion time of plan $\pi$, \sys uses the cache and load state in the cluster view.
Here, $\mathcal H_w$ denotes the set of segments cached at worker $w$, and $Q_w$ is its total outstanding compute load, including scatter and combine prefill.
$Q_w$ is measured in tokens and excludes the current request.
Let $\mathcal S_w(\pi)=\{i\in\mathcal S\mid a(i)=w\}$ denote the segments assigned to worker $w$.
For cache hits $\mathcal S_w(\pi)\cap\mathcal H_w$, \sys reuses the KV already stored at that worker; for cache misses $\mathcal S_w(\pi)\setminus\mathcal H_w$, it performs scatter prefill there.

We estimate the prefill completion time of plan $\pi$ as:
\begin{equation}
    \label{eq:prefill-completion-time}
    \widehat T(\pi)=
    \max_{w\in\mathcal W}
    \left[
        Q_w(\pi)+P_w(\pi)+D_w(w\!\to c;\pi)
    \right]
    +C.
\end{equation}
The estimate sums queueing $Q_w(\pi)$, computation $P_w(\pi)$, and transfer time $D_w(w\!\to c;\pi)$ at each worker, takes the maximum across workers as the estimated KV-ready time, and adds the request's combine prefill work $C$.
In this model, matched prefix KV is also represented as cached segments and included in the transfer cost.
All terms are normalized to equivalent prefill tokens:
\begin{equation}
    \begin{aligned}
        Q_w(\pi)
        &=\begin{cases}
              Q_w, & \mathcal S_w(\pi)\setminus\mathcal H_w\ne\varnothing,\\
              0, & \text{otherwise},
        \end{cases}\\[4pt]
        P_w(\pi)
        &=\sum_{i\in\mathcal S_w(\pi)\setminus\mathcal H_w}P_i,\\[4pt]
        D_w(w\!\to c;\pi)
        &=d(w\!\to c)\sum_{i\in\mathcal S_w(\pi)}P_i.
    \end{aligned}
\end{equation}
Here, $P_i$ is the token count of segment $i$.
The per-worker queueing and scatter-computation terms arise only when a worker is assigned a cache miss.
The transfer term covers both cache hits and newly computed KV.
The coefficient $d(w\!\to c)$ is obtained through offline profiling: it is the per-token KV transfer time from worker $w$ to worker $c$, normalized by the per-token prefill time, with $d(c\!\to c)=0$.
The request's combine prefill workload $C$, which includes computation of the remaining content and necessary PIC recomputation, is constant across candidate plans.
Combine prefill receives the highest scheduling priority once its required KV is ready, so the estimate omits its queueing delay.

\phm{Greedy Plan Generation.}
Using this prefill completion time estimate, \sys greedily optimizes the execution plan to minimize completion time (Algorithm~\ref{alg:prefill-planning}).

The algorithm first enumerates available combine homes $c$ and constructs an initial plan $\pi$ for each: cache hits are assigned to the source workers, and all cold PIC segments are assigned to $c$ (lines~\ref{line:prefill-combine}--\ref{line:prefill-init}).
This initial assignment preserves cluster-wide cache reuse while avoiding KV return traffic from cold PIC-segment computation.

For each cold PIC segment in decreasing order of length, the algorithm constructs a candidate plan $\pi'$ by reassigning it to the worker $w\ne c$ minimizing $Q_w+P_w(\pi)$ (lines~\ref{line:prefill-scatter}--\ref{line:prefill-candidate}).
The search holds $Q_w$ fixed, while updating $P_w(\pi)$ after each accepted move.
Load guides candidate selection, but a move is accepted only if $\widehat T(\pi')<\widehat T(\pi)$ (lines~\ref{line:prefill-accept}--\ref{line:prefill-update}).
Otherwise, the plan remains unchanged and the search proceeds to the next segment.

After searching all candidate combine homes, it returns a plan with the lowest estimated completion time (lines~\ref{line:prefill-best}--\ref{line:prefill-return}).

\begin{figure*}[t]
    \centering
    \includegraphics[width=0.95\linewidth]{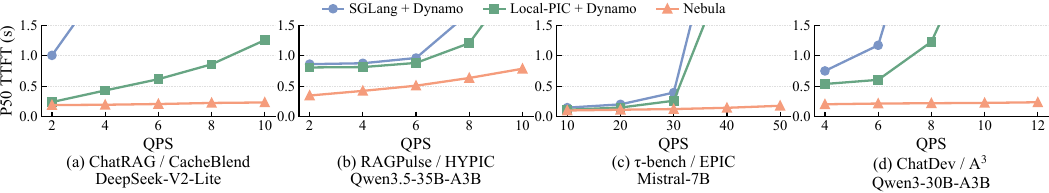}
    \caption{End-to-end median TTFT versus QPS: (a) ChatRAG with DeepSeek-V2-Lite and CacheBlend, (b) RAGPulse with Qwen3.5-35B-A3B and HYPIC, (c) $\tau$-bench with Mistral-7B and EPIC, and (d) ChatDev with Qwen3-30B-A3B and A$^3$. }
    \label{fig:eval-e2e-ttft}
    \par\medskip
    \includegraphics[width=0.95\linewidth]{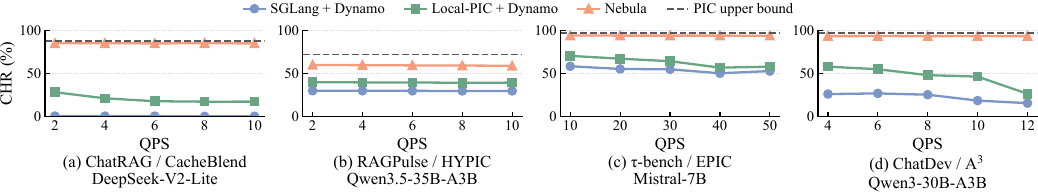}
    \caption{End-to-end cache hit ratio versus QPS, with the same configurations as Figure~\ref{fig:eval-e2e-ttft}. }
    \label{fig:eval-e2e-cache-hit}
    \par\medskip
    \includegraphics[width=0.95\linewidth]{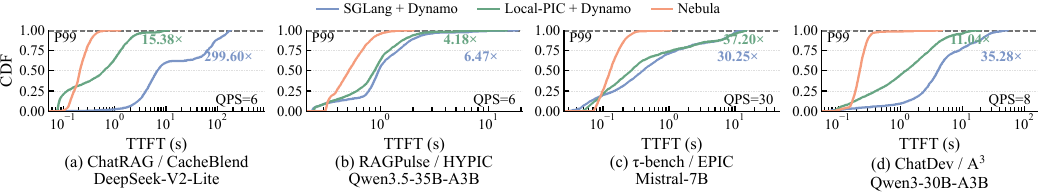}
    \caption{End-to-end TTFT CDFs, with the same configurations as Figure~\ref{fig:eval-e2e-ttft}.}
    \label{fig:eval-e2e-ttft-cdf}
    \vspace{-2ex}
\end{figure*}

\section{Implementation}
\label{sec:eval-implementation}

We implement \sys on SGLang~\cite{zheng2023efficiently} and Dynamo~\cite{dynamo_router}, reusing SGLang's radix tree and tiered storage infrastructure and Dynamo's cluster orchestration.
On the workers, we extend SGLang's forward execution path to support CacheBlend~\cite{yao2025cacheblend}, A$^3$~\cite{zhou2025a3}, EPIC~\cite{hu2025epic} and HYPIC~\cite{liu2026hypic}, and adapt its storage pool allocation and indexing to manage PIC segments across storage tiers.
On the router, we extend Dynamo's cluster directory to support cache routing and prefill scheduling for distributed PIC (\S\ref{sec:architecture}).
We use NIXL~\cite{nvidia2026nixl} for efficient point-to-point transfers of PIC between workers across storage tiers.
Our implementation comprises approximately 19K lines of code.

\section{Evaluation}
\label{sec:eval}

\subsection{Setup}
\label{sec:eval-setup}

\phm{Hardware.}
We run our experiments on two identical servers, each equipped with four NVLink-connected NVIDIA H20-3e GPUs, two 48-core Intel Xeon Platinum 8575C CPUs, 1\,TiB of host memory, and four 200\,Gbps RDMA-capable Ethernet ports.

\phm{Models.}
We evaluate four models: DeepSeek-V2-Lite~\cite{deepseekai2024deepseekv2}, Qwen3.5-35B-A3B~\cite{qwen3.5}, Mistral-7B~\cite{mistralai2023mistralv02}, and Qwen3-30B-A3B~\cite{qwen3technicalreport}.
All models use seven prefill workers and one decode worker (7P1D), with one GPU per worker (TP1).
We run DeepSeek-V2-Lite with CacheBlend, Qwen3.5-35B-A3B with HYPIC, Mistral-7B with EPIC, and Qwen3-30B-A3B with A$^3$.

\phm{Workloads.}
We use four public RAG and agent workloads: ChatRAG-QuAC~\cite{liu2024chatqa,choi2018quac} for conversational RAG, RAGPulse~\cite{wang2025ragpulse} for RAG serving, $\tau$-bench-Retail~\cite{yao2025tau} for tool-using agents, and ChatDev~\cite{qian2024chatdev} for multi-agent software development.
Table~\ref{tab:pic-opportunity} summarizes the theoretical cache-reuse breakdown for these workloads, while Table~\ref{tab:segment-distribution} reports the distributions of segment counts and lengths.
We preserve the original request order and replay each trace with Poisson arrivals at varying rates.
Since PIC targets TTFT reduction, we limit each request to one output token while retaining one decode worker to preserve prefill--decode disaggregation deployment.

\phm{Methods.}
We compare three configurations:
(i) \textbf{SGLang + Dynamo} combines SGLang~\cite{zheng2023efficiently} workers with Dynamo's prefix-cache-aware router~\cite{dynamo_router}, representing a common production deployment;
(ii) \textbf{Local-PIC + Dynamo} is the direct-composition strawman described in \S\ref{sec:distributed-pic}, deploying PIC-enabled workers behind Dynamo without remote cache reuse or parallel prefill of PIC segments; and
(iii) \textbf{\sys} integrates our three system designs to support remote PIC reuse and parallel prefill.
Component experiments vary the mechanism under study while holding the remaining configuration fixed.
Unless otherwise specified, each configuration runs from a fresh cluster without warmup.

\subsection{End-to-end performance}
\label{sec:eval-e2e}

We evaluate \sys's end-to-end performance using the four real-world workloads and four models described in~\S\ref{sec:eval-setup}.

\phm{Median and tail TTFT.}
Figure~\ref{fig:eval-e2e-ttft} shows that \sys achieves the lowest median TTFT across all measured settings.
Across these settings, \sys reduces median TTFT by an average of 61.4\% and 76.7\% relative to Local-PIC + Dynamo and SGLang + Dynamo, respectively.
At the highest tested rate for each workload, its median TTFT is 0.24, 0.79, 0.18, and 0.24\,s for ChatRAG, RAGPulse, $\tau$-bench-Retail, and ChatDev, respectively, compared with 1.26, 2.39, 3.64, and 9.57\,s for Local-PIC + Dynamo and 75.73, 2.71, 8.79, and 8.46\,s for SGLang + Dynamo.
The results demonstrate the end-to-end benefits of jointly supporting remote cache reuse and parallel prefill beyond worker-local PIC alone.

We also report the TTFT CDF in Figure~\ref{fig:eval-e2e-ttft-cdf} at 6, 6, 30, and 8 QPS for ChatRAG, RAGPulse, $\tau$-bench-Retail, and ChatDev, respectively.
At P99, \sys reduces TTFT by 93.5\%, 76.1\%, 97.3\%, and 90.9\% relative to Local-PIC + Dynamo on these workloads, respectively.
Relative to SGLang + Dynamo, the reductions range from 84.6\% to 99.7\%.

\phm{TTFT-constrained goodput.}
We measure goodput as the number of successful requests meeting the TTFT SLO divided by the time from the first request's start to the last request's completion.
With a 1\,s TTFT SLO, \sys achieves 11.66 requests/s on ChatRAG at 12 QPS, compared with 1.68 requests/s for Local-PIC + Dynamo.

\phm{Cache hit ratio.}
We measure Cache Hit Ratio (CHR) as the total number of input tokens served from cache divided by the total number of input tokens.
Figure~\ref{fig:eval-e2e-cache-hit} shows that \sys achieves the highest CHR across all measured settings.
At the highest tested rates, its CHRs are 84.82\%, 58.99\%, 93.67\%, and 93.17\% for ChatRAG, RAGPulse, $\tau$-bench-Retail, and ChatDev, respectively, exceeding Local-PIC + Dynamo by 19.48--67.23 percentage points.

Compared with SGLang + Dynamo, \sys recovers 84.7\%, 42.1\%, 86.3\%, and 91.2\% of the tokens missed by prefix caching on ChatRAG, RAGPulse, $\tau$-bench-Retail, and ChatDev, respectively.

\subsection{View Synchronization}
\label{sec:eval-view}

\begin{figure}[t]
    \centering
    \subfloat[RAGPulse.\label{fig:eval-view-ragpulse}]{%
        \includegraphics[width=0.95\linewidth]{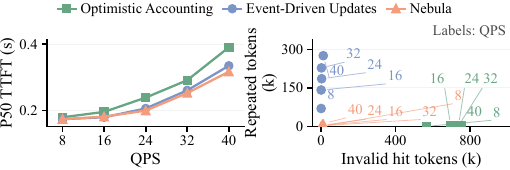}%
    }\\
    \subfloat[$\tau$-bench-Retail.\label{fig:eval-view-tau-retail}]{%
        \includegraphics[width=0.95\linewidth]{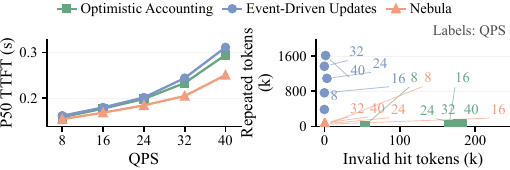}%
    }
    \caption{
        View synchronization at different request rates. The left plots show median TTFT, while the right plots compare invalid hit and repeated computation token counts.
    }
    \label{fig:eval-view}
    \vspace{-2ex}
\end{figure}

\phm{Invalid hits and repeated computation.}
We compare \sys with Optimistic Accounting and Event-Driven Updates described in~\S\ref{sec:cluster-view-challenge}, varying only the PIC directory maintenance mechanism.
Using Qwen3.5-35B-A3B, we replay the RAGPulse and $\tau$-bench-Retail traces at different request rates.
\sys uses a 5\,s TTL for unconfirmed entries.

Figure~\ref{fig:eval-view} shows that \sys achieves the lowest median TTFT on both workloads.
At 40 QPS, it reduces median TTFT by 18.9\% on RAGPulse and 14.9\% on $\tau$-bench-Retail relative to Optimistic Accounting.
Relative to Event-Driven Updates, the reductions are 5.4\% and 19.5\%, respectively.

At 40 QPS, \sys reduces the number of invalid hit tokens by 98.4\% on RAGPulse and 100\% on $\tau$-bench-Retail relative to Optimistic Accounting.
Relative to Event-Driven Updates, it reduces the number of repeated computation tokens by 96.1\% and 95.1\%, respectively.

\subsection{Cache Management}
\label{sec:eval-cache}

\begin{figure}[t]
    \centering
    \includegraphics[width=0.95\linewidth]{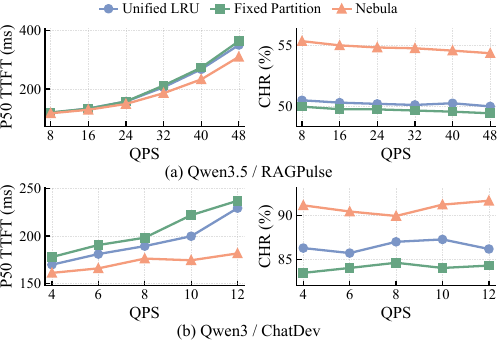}
    \caption{Cache management policies versus QPS for (a) Qwen3.5-35B-A3B on RAGPulse and (b) Qwen3-30B-A3B on ChatDev.}
    \label{fig:eval-cache}
    \vspace{-2ex}
\end{figure}

\phm{Cache capacity management variants.}
We compare \sys with Unified LRU and Fixed Partition from \S\ref{sec:cache-management-challenge}, holding view synchronization and prefill scheduling fixed.
We run Qwen3.5-35B-A3B on RAGPulse and Qwen3-30B-A3B on ChatDev at different request rates, both on a 7P1D cluster with TP1.
Fixed Partition allocates half of the cache capacity to PC and half to PIC.
In this experiment, we use the first 800 requests as warmup and report metrics over the remaining requests.

Figure~\ref{fig:eval-cache} shows that \sys achieves the lowest TTFT and highest CHR.
On RAGPulse at 48 QPS, \sys increases CHR by 4.32 and 4.88 percentage points and reduces median TTFT by 11.5\% and 14.7\% relative to Unified LRU and Fixed Partition, respectively.
On ChatDev at 12 QPS, the corresponding CHR gains are 5.48 and 7.37 percentage points, with median TTFT reductions of 20.8\% and 23.4\%.

\begin{figure}[t]
    \centering
    \includegraphics[width=0.95\linewidth]{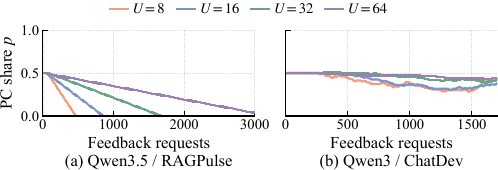}
    \caption{Trajectory of the PC target soft share $p$ under different update windows $U$ for (a) Qwen3.5-35B-A3B on RAGPulse and (b) Qwen3-30B-A3B on ChatDev.}
    \label{fig:eval-cache-p}
    \vspace{-2ex}
\end{figure}

\begin{figure}[t]
    \centering
    \includegraphics[width=0.95\linewidth]{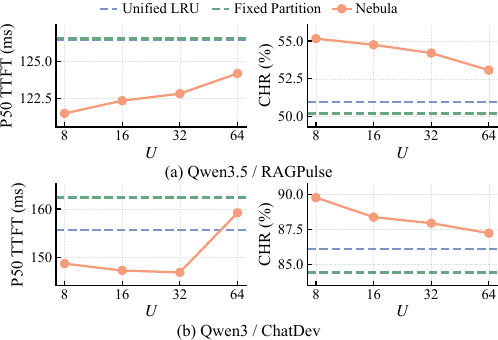}
    \caption{Effect of the update window $U$: median TTFT and CHR for (a) Qwen3.5-35B-A3B on RAGPulse and (b) Qwen3-30B-A3B on ChatDev. Dashed lines show the two baselines, which do not depend on $U$.}
    \label{fig:eval-cache-window}
    \vspace{-2ex}
\end{figure}

\phm{Update-window sensitivity and capacity-share trajectory.}
We next study the update-window sensitivity and how the update window affects the shares of PC and PIC.
We run Qwen3.5-35B-A3B on RAGPulse at 12 QPS and Qwen3-30B-A3B on ChatDev at 6 QPS, varying the update window $U$ over 8, 16, 32, and 64 requests.

Figure~\ref{fig:eval-cache-p} shows how the PC target share $p$ evolves under different update windows.
On RAGPulse, $p$ approaches zero at the maximum adjustment rate, consistent with its preference for PIC and few PC hits.
On ChatDev, $p$ decreases and then rises as feedback changes, indicating that the relative demand for PC and PIC varies over time.

Figure~\ref{fig:eval-cache-window} shows that $U=8$ gives the lowest measured median TTFT on RAGPulse (121.46\,ms), while $U=32$ gives the lowest on ChatDev (146.94\,ms).
The $U=16$ results are 122.33\,ms and 147.32\,ms, respectively, achieving median TTFT within 1\,ms of the best tested window on both workloads.
Considering the update frequency as well, we use $U=16$ by default.

\subsection{Prefill Scheduling}
\label{sec:eval-scheduling}

\begin{figure}[t]
    \centering
    \subfloat[RAGPulse.\label{fig:eval-scheduling-ragpulse}]{%
        \includegraphics[width=0.95\linewidth]{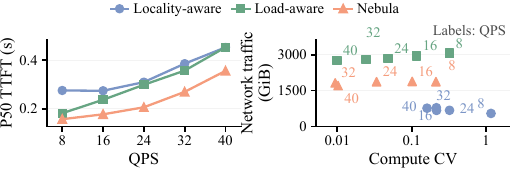}%
    }\\
    \subfloat[$\tau$-bench-Retail.\label{fig:eval-scheduling-tau-retail}]{%
        \includegraphics[width=0.95\linewidth]{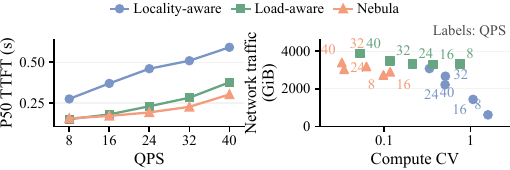}%
    }
    \caption{
        Prefill scheduling at different request rates. The left plots show median TTFT, while the right plots compare total NIXL TX payload and compute CV.
    }
    \label{fig:eval-scheduling}
    \vspace{-2ex}
\end{figure}

\phm{Traffic and Load Balance.}
We compare \sys with the Locality-aware and Load-aware strawmen from \S\ref{sec:scheduling-challenge}.
We deploy Qwen3.5-35B-A3B and replay the RAGPulse and $\tau$-bench-Retail traces at different request rates.
All schedulers support remote PIC reuse.

Figure~\ref{fig:eval-scheduling} shows that \sys achieves the lowest median TTFT on both workloads at 24--40 QPS.
At 40 QPS, it reduces median TTFT by 21.4\% and 48.9\% relative to Locality-aware, and by 21.4\% and 19.8\% relative to Load-aware, on RAGPulse and $\tau$-bench-Retail, respectively.

We measure total NIXL TX payload and the population CV of cumulative prefill tokens across prefill workers.
At 40 QPS, \sys reduces traffic by 38.1\% and 12.3\% relative to Load-aware, and compute CV by 93.6\% and 90.3\% relative to Locality-aware, on RAGPulse and $\tau$-bench-Retail, respectively.
At this rate, Locality-aware transfers less data than \sys but has higher compute CV on both workloads.

\begin{figure}[t]
    \centering
    \includegraphics[width=0.95\linewidth]{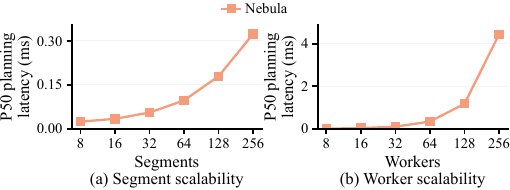}
    \caption{Scheduling overhead of \sys. (a) Varying the number of segments with 16 workers. (b) Varying the number of workers with 16 segments.}
    \label{fig:eval-scheduling-overhead}
    \vspace{-2ex}
\end{figure}

\phm{Scheduling overhead.}
We measure the latency of greedy plan generation (\S\ref{sec:scheduling-design}) on one pinned vCPU.
We uniformly sample segment lengths from 128--2,048 tokens and outstanding worker loads from 0--8,192 tokens, with a 50\% segment cache hit ratio and each cache hit placed on a uniformly selected worker.
For each setting, we measure 1,024 planning calls after 64 warmup calls.
Figure~\ref{fig:eval-scheduling-overhead} shows that median planning latency is 0.32\,ms with 256 segments and 16 workers, and 4.43\,ms with 16 segments and 256 workers.

    \section{Conclusion}
\label{sec:conclusion}

\sys is the first distributed LLM serving framework to treat PIC segments as first-class citizens, enabling remote cache reuse and parallel prefill.
It combines optimistic accounting with event-driven updates to maintain a timely and truthful cluster view, adapts PC/PIC capacity shares using cluster-wide ghost feedback, and selects prefill placement and parallelism through a unified completion-time estimate.
Across four models and four real-world RAG and agentic workloads, \sys reduces median TTFT by an average of 61.4\% compared with the strongest existing baseline and recovers an average of 76.1\% of the tokens missed by prefix caching.

\end{document}